\PassOptionsToPackage{unicode}{hyperref}
\PassOptionsToPackage{hyphens}{url}
\PassOptionsToPackage{dvipsnames,svgnames,x11names}{xcolor}
\documentclass[
  11pt,
  a4paper,
]{article}
\usepackage{amsmath,amssymb}
\usepackage{setspace}
\usepackage{iftex}
\ifPDFTeX
  \usepackage[T1]{fontenc}
  \usepackage[utf8]{inputenc}
  \usepackage{textcomp} 
\else 
  \usepackage{unicode-math} 
  \defaultfontfeatures{Scale=MatchLowercase}
  \defaultfontfeatures[\rmfamily]{Ligatures=TeX,Scale=1}
\fi
\usepackage[]{mathpazo}
\ifPDFTeX\else
\fi
\IfFileExists{upquote.sty}{\usepackage{upquote}}{}
\IfFileExists{microtype.sty}{
  \usepackage[]{microtype}
  \UseMicrotypeSet[protrusion]{basicmath} 
}{}
\makeatletter
\@ifundefined{KOMAClassName}{
  \IfFileExists{parskip.sty}{%
    \usepackage{parskip}
  }{
    \setlength{\parindent}{0pt}
    \setlength{\parskip}{6pt plus 2pt minus 1pt}}
}{
  \KOMAoptions{parskip=half}}
\makeatother
\usepackage{xcolor}
\usepackage[margin=1in]{geometry}
\usepackage{longtable,booktabs,array}
\usepackage{calc} 
\usepackage{etoolbox}
\makeatletter
\patchcmd\longtable{\par}{\if@noskipsec\mbox{}\fi\par}{}{}
\makeatother
\IfFileExists{footnotehyper.sty}{\usepackage{footnotehyper}}{\usepackage{footnote}}
\makesavenoteenv{longtable}
\usepackage{graphicx}
\makeatletter
\def\maxwidth{\ifdim\Gin@nat@width>\linewidth\linewidth\else\Gin@nat@width\fi}
\def\maxheight{\ifdim\Gin@nat@height>\textheight\textheight\else\Gin@nat@height\fi}
\makeatother
\setkeys{Gin}{width=\maxwidth,height=\maxheight,keepaspectratio}
\makeatletter
\def\fps@figure{htbp}
\makeatother
\NewDocumentCommand\citeproctext{}{}
\NewDocumentCommand\citeproc{mm}{%
  \begingroup\def\citeproctext{#2}\cite{#1}\endgroup}
\makeatletter
 \let\@cite@ofmt\@firstofone
 \def\@biblabel#1{}
 \def\@cite#1#2{{#1\if@tempswa , #2\fi}}
\makeatother
\newlength{\cslhangindent}
\newlength{\csllabelwidth}
\newenvironment{CSLReferences}[2] 
 {\begin{list}{}{%
  \setlength{\itemindent}{0pt}
  \setlength{\leftmargin}{0pt}
  \setlength{\parsep}{0pt}
  \ifodd #1
   \setlength{\leftmargin}{\cslhangindent}
   \setlength{\itemindent}{-1\cslhangindent}
  \fi
  \setlength{\itemsep}{#2\baselineskip}}}
 {\end{list}}
\usepackage{calc}

\usepackage{amsmath} 
\usepackage{float} 

\usepackage{lscape}
\newcommand{\blandscape}{\begin{landscape}}
\newcommand{\elandscape}{\end{landscape}}

\newcommand{\bleft}{\begin{flushleft}}
\newcommand{\eleft}{\end{flushleft}}

\ifLuaTeX
  \usepackage{selnolig}  
\fi
\usepackage{bookmark}
\IfFileExists{xurl.sty}{\usepackage{xurl}}{} 
\hypersetup{
  pdftitle={Estimating invasive rodent abundance using removal data and hierarchical models},
  pdfauthor={Olivier Gimenez1*},
  colorlinks=true,
  linkcolor={RoyalBlue},
  filecolor={Maroon},
  citecolor={Blue},
  urlcolor={RoyalBlue},
  pdfcreator={LaTeX via pandoc}}

\title{Estimating invasive rodent abundance using removal data and hierarchical models}
\author{Olivier Gimenez\textsuperscript{1}*}
\date{2025-04-19}

\begin{document}
\maketitle

\setstretch{2}

\small

\textsuperscript{1} CEFE, Univ Montpellier, CNRS, EPHE, IRD, Montpellier, France

\texttt{*} Corresponding author: \href{mailto:olivier.gimenez@cefe.cnrs.fr}{\nolinkurl{olivier.gimenez@cefe.cnrs.fr}}

\normalsize

\vspace{1cm}
\hrule

Invasive rodents pose significant ecological, economic, and public health challenges. Robust methods are needed for estimating population abundance to guide effective management. Traditional methods such as capture-recapture are often impractical for invasive species due to ethical, legal and logistical constraints. Here, I showcase the application of hierarchical multinomial N-mixture models for estimating the abundance of invasive rodents using removal data. First, I perform a simulation study which demonstrates minimal bias, as well as good precision and reliable coverage of confidence intervals across a range of sampling scenarios. I also illustrate the consequences of violating the population closure assumption, showing how between-occasion dynamics can bias inference. Second, I analyze removal data for two invasive rodent species, namely coypus (\emph{Myocastor coypus}) in France and muskrats (\emph{Ondatra zibethicus}) in the Netherlands. Using hierarchical multinomial N-mixture models, I examine the effects of temperature on abundance while accounting for imperfect and time-varying capture probabilities. I also show how to accommodate spatial variability using random effects, quantify uncertainty in parameter estimates, and account for violations of closure by fitting an open-population model to multi-year data. Overall, I hope to demonstrate the flexibility and utility of hierarchical models in invasive species management.
\vspace{3mm}

\hrule
\vspace{5mm}


\newpage

\section{Introduction}\label{introduction}

Invasive species are a significant global issue, with wide-ranging impacts on ecosystems, economies, and public health (\citeproc{ref-Petr2020}{Pyšek et al. 2020}, \citeproc{ref-Roy2024}{Roy et al. 2024}). Among these, the financial, epidemiological, social, and ecological costs associated with invasive rodents are substantial, as they damage infrastructures, degrade agricultural systems, and act as reservoirs for zoonotic diseases (\citeproc{ref-Diagne2023}{Diagne et al. 2023}).

Effective management of invasive species requires the estimation of population abundance for guiding control efforts and evaluating the success of eradication or regulation programs (\citeproc{ref-Williams2002}{Williams et al. 2002}, \citeproc{ref-Thompson2021}{Thompson et al. 2021}). However, the challenge in estimating animal abundance is that, because of imperfect detection, individuals are not always observed even when present (\citeproc{ref-Borchers2002}{Borchers et al. 2002}, \citeproc{ref-Seber2023}{Seber and Schofield 2023}). Ignoring imperfect detection leads to biased estimates of population abundance (\citeproc{ref-Kery2008}{Kéry and Schmidt 2008}). To account for imperfect detection, capture-recapture methods are usually used to correct observed counts (\citeproc{ref-Mccrea2015}{McCrea and Morgan 2015}). Yet, for invasive species, capture-recapture is often impractical, as ethical and management concerns typically prevent the release of captured animals.

An alternative approach involves the use of removal methods (\citeproc{ref-Rodriguez2021}{Rodriguez de Rivera and McCrea 2021}) in which individuals are captured and permanently removed from the study area during successive sampling occasions. This process leads to a decrease in the expected number of captures by a consistent proportion over time (rather than by a fixed amount decline), which informs on the total abundance as the initial population determines how quickly the number of individuals available for capture diminishes.

While standard removal methods are well-established (\citeproc{ref-Moran1951}{Moran 1951}, \citeproc{ref-Zippin1956}{Zippin 1956}, \citeproc{ref-Zippin1958}{1958}, \citeproc{ref-Rodriguez2021}{Rodriguez de Rivera and McCrea 2021}), recent advances in population ecology remain underutilized in the context of invasive species. Hierarchical models, in particular, have gained traction (\citeproc{ref-RD2008}{Royle and Dorazio 2008}, \citeproc{ref-KR2015}{Kéry and Royle 2015}) due to their ability to: (i) explicitly separate biological processes of interest (e.g., population dynamics) from observation processes (e.g., imperfect detection), thus enabling more accurate modeling; (ii) incorporate environmental, spatial, or temporal covariates at multiple levels, allowing exploration of how various factors influence ecological processes; and (iii) share information across groups (e.g., years) by modeling parameters hierarchically with random effects, which improves estimates for groups with fewer data.

In this paper, I showcase the application of a hierarchical formulation of removal models, the multinomial N-mixture model (\citeproc{ref-Dorazio2005}{Dorazio et al. 2005}), to estimate the abundance of rodents in Europe. In this study, I focus on the coypu (\emph{Myocastor coypus}) in France and the muskrat (\emph{Ondatra zibethicus}) in the Netherlands. Both species are semi-aquatic rodents introduced to Europe in the early 20th century following escapes or releases from fur farms. The coypu, native to South America, has formed widespread invasive populations in France (\citeproc{ref-Bonnet2023}{Bonnet et al. 2023}), where it causes significant damage to infrastructure and crops. Additionally, it serves as a healthy carrier of leptospirosis, a zoonotic disease with potentially serious consequences. Similarly, the muskrat, native to North America, has established extensive populations in the Netherlands. By burrowing into riverbanks, dykes, and dams, muskrats compromise the integrity of these structures, posing a threat to public safety (\citeproc{ref-vanloon2017}{Loon et al. 2017}). Both species are also widespread in other European countries; updated distribution maps are available via the European Alien Species Information Network (EASIN) platform (\url{https://easin.jrc.ec.europa.eu/spexplorer/search/}).

Using removal data, I demonstrate the application of the multinomial N-mixture model to estimate the abundance of rodent populations. First, I conduct a simulation study to evaluate the model's performance under varying numbers of sampling sites and sampling occasions. Second, I present a case study on a coypu population in France to illustrate the hierarchical structure of the multinomial N-mixture model, demonstrating how covariates can be incorporated to account for variations in abundance and capture probabilities. Third, I use a case study on muskrats in the Netherlands to illustrate the integration of random effects within the model and demonstrate how to relax the closure assumption. To facilitate reproducibility, I provide the accompanying code and data, aiming to promote the broader adoption of removal models in the study of biological invasions.

\section{Methods}\label{methods}

\subsection{Multinomial N-mixture model}\label{multinomial-n-mixture-model}

Think of a dice with six sides. The dice has a 1 in 6 chance of landing on face 1, the same for face 2, and so on. If I roll the dice 30 times, I would expect, on average over many repetitions of this experiment, to get face 1 five times, face 2 five times, and so on. You can test this in \texttt{R} by running the command \texttt{rmultinom(n\ =\ 1,\ size\ =\ 30,\ prob\ =\ c(1/6,\ 1/6,\ 1/6,\ 1/6,\ 1/6,\ 1/6))} repeatedly. In this experiment, \(y_1\), the number of 1s, \(y_2\), the number of 2s, \(\ldots\), and \(y_6\), the number of 6s, follows a multinomial distribution with parameters the number of rolls (30) and probabilities \((1/6, 1/6, ..., 1/6)\).

Now think of a removal campaign conducted over 3 months. We record the number of rodents \(y_1\) captured in month 1, \(y_2\) in month 2, \(y_3\) in month 3, and let \(y_4\) represent the number of rodents never captured. Let \(p\) be the probability of capturing a rodent in a given month. The probability of capturing a rodent in the first month is \(\pi_1 = p\). The probability of capturing a rodent in the second month is \(\pi_2 = (1-p)p\) the probability of not capturing it in the first month \((1 - p)\) multiplied by the probability of capturing it in the second month \(p\). The probability of capturing a rodent in the third month is \(\pi_3 = (1-p)(1-p)p\), the probability of not capturing it in the first and second months, \((1 - p)(1 - p)\), multiplied by the probability of capturing it in the third month, \(p\). Finally, the probability of never being captured is \(\pi_4 = 1 - (\pi_1 + \pi_2 + \pi_3)\) the complement of the probability of being captured in the first, second, or third month. If we assume that \(N\) represents the abundance, then we have that the vector of counts \((y_1, y_2, y_3, y_4)\) follows a multinomial distribution with parameters \(N\) and probabilities \((\pi_1,\pi_2,\pi_3,\pi_4)\). This is the observation process. In general, we assume that \(N\) follows a Poisson distribution with parameter the expected number of rodents denoted \(\lambda\). This is the state or ecological process. And there you have it, the multinomial N-mixture model for a removal experiment, which is similar to throwing a dice \(N\) times and the \(\pi's\) give the probabilities that I get a given face of that dice. Unlike a fair die, however, the probabilities in a removal experiment are not equal; they reflect varying detection probabilities over time, which depend on factors like effort, animal behavior, or environmental conditions. Also, in general, we monitor rodents in several populations \(i = 1,\ldots,S\) and we need to estimate local abundance \(N_i\). To do so, Dorazio et al. (\citeproc{ref-Dorazio2005}{2005}) extended multinomial N-mixture models to account for spatial variation in abundance and/or capture, and showed that abundance estimates had similar or better precision than those obtained from analyzing removal data for each population separately.

Parameters \(N\), \(p\), and \(\lambda\) are unknown and need to be estimated. In a frequentist framework, marginalization is performed by summing over all possible values of \(N\) (\citeproc{ref-Dorazio2005}{Dorazio et al. 2005}). In a Bayesian framework, all these parameters are estimated directly, which simplifies the process (\citeproc{ref-RD2006}{Royle and Dorazio 2006}). Both parameters, \(\lambda\) and \(p\), can be modeled as functions of explanatory spatial and temporal variables, in the spirit of generalized linear models, and Poisson (with a log link function) or logistic regressions (with a logit link function) for example.

To evaluate model adequacy, I used standard goodness-of-fit procedures adapted to both frequentist and Bayesian frameworks. In the frequentist framework, we apply a parametric bootstrap approach: we generate a large number of replicate datasets from the maximum likelihood estimates, refit the model to each replicate, and compute diagnostic statistics such as the Freeman--Tukey statistic. If the resulting bootstrap p-values fall within a non-extreme range, this indicates no evidence of lack of fit. In the Bayesian framework, I assessed model adequacy using posterior predictive checks based on Bayesian p-values. At each MCMC iteration, a replicate dataset is drawn from the joint posterior distribution, and the Freeman--Tukey discrepancy is computed for both the observed and replicate data. Given the conditional multinomial structure of the model, which separates the observation process from the abundance process, I calculated two discrepancy measures: one for the detection histories and another for the total counts observed. Bayesian p-values near 0.5 (and away from 0 or 1) indicate no evidence of systematic lack of fit.

For a detailed description of the multinomial mixture model, I warmly recommend chapter 7 in Kéry and Royle (\citeproc{ref-KR2015}{2015}).

\section{Simulation study}\label{simulation-study}

I conducted a simulation study to evaluate the model's performance by examining parameter bias under varying numbers of sampling sites and sampling occasions. I simulated removal data over 1, 5, 10 and 50 sites using a Poisson distribution with expected number of animals \(\lambda\) between 10 and 100 (20 values) for the ecological process. I simulated the observation process with a capture probability \(p\) varying between 0.3 and 0.9 (20 values) across 3, 5 and 10 occasions per site. In total, I considered 4800 scenarios. I fitted the multinomial N-mixture model to the simulated data within the frequentist framework using function \texttt{multinomPois()} in the \texttt{R} package \texttt{unmarked} (\citeproc{ref-Kellner_2023}{Kellner et al. 2023}), and I repeated this procedure 100 times. Eventually, I calculated the relative bias, root mean square error (RMSE), and coverage of the 95\% confidence interval for each parameter.

To assess the effect of violating the closure assumption, I implemented an additional set of simulations in which the population could change between sampling occasions. Specifically, individuals staying in the population with probability 0.8, and new individuals arrive according to a Poisson process with mean 1. Apart from these between-occasion dynamics, all other aspects of the simulation setup remained the same. This setup breaks the closure assumption in two ways. Some individuals leave the population between sampling occasions, violating the assumption that declines in abundance are due to removal alone; this can bias detection probability and abundance estimates. New individuals enter between sampling occasions, inflating the pool of animals available for detection and leading to overestimation of abundance. Since I deliberately fit a closed model to data from an open process, any resulting bias directly reflects the impact of violating closure. While this simulation focuses on geographic closure, the same logic applies to demographic closure, where the stay and arrivals parameters correspond to survival and recruitment processes.

Note that I used a frequentist implementation for the simulation study to reduce computation time given the large number of scenarios. The model structure remains hierarchical, as in the Bayesian case studies, and both inferential approaches would yield similar results. The aim was to assess model performance across ecological and sampling conditions, not to compare statistical paradigms.

\section{Case studies}\label{case-studies}

In this section, I analyzed removal data from two rodent species: coypus in France and muskrats in the Netherlands. With these case studies, I aimed at illustrating specific features of hierarchical multinomial N-mixture models. For both species, I explored the potential effect of temperature on abundance (e.g., \citeproc{ref-Gosling1981}{Gosling 1981}, \citeproc{ref-Simpson1993}{Simpson and Boutin 1993}). A comprehensive analysis of the ecological factors influencing population dynamics was beyond the scope of this work and will be addressed in future studies.

\subsection{Coypus in France}\label{coypus-in-france}

Removal data on coypus were collected from annual control operations conducted since 2015 in several cities within the Hérault department, located in the Occitanie region of southern France. These operations are carried out year-round, with the exception of July and August. Coypus are trapped using cages by a network of volunteers coordinated by the Syndicat Mixte du Bassin de l'Or and the Fédération Départementale des Chasseurs de l'Hérault (\url{https://etang-de-l-or.com/lutte-ragondins/}). For this study, I focus on data from 2022, specifically from sampling occasions in February, March, and April. The data, covering \(S = 6\) cities, are summarized in Table \ref{tab:coypus}. I fitted a model where the expected number of coypus was modeled as a function of temperature, while the capture probability was allowed to vary by month.

\begin{table}[ht]
\centering
\begin{tabular}{lcccc}
\hline
\textbf{} & \textbf{Removed in} & \textbf{Removed in} & \textbf{Removed in} & \textbf{Averaged} \\
\textbf{City} & \textbf{February} & \textbf{March} & \textbf{April} & \textbf{temperature}\\
\hline
Candillargues & 18  & 12    & 38 & 9.5 \\
Lansargues    & 15  & 17    & 75 & 8.8 \\
Mauguio       & 20  & 9 & 6 & 9.2 \\
Saint-Nazaire-de-Pézan & 169    & 41    & 15 & 9.3 \\
Saint-Just    & 85  & 61    & 77 & 9.2 \\
Valergues     & 0 & 1   & 3 & 9.4 \\
\hline
\end{tabular}
\caption{Number of invasive coypus removed monthly and the average 3-month temperature across several cities in the Hérault department, France, in 2022.}
\label{tab:coypus}
\end{table}

A key assumption of the multinomial N-mixture model is that abundance follows a Poisson distribution, which implies equal mean and variance. When this assumption is violated - i.e., in the presence of overdispersion - a common and effective solution is to replace the Poisson with a negative binomial distribution. I illustrate how to fit such an overdispersed model using the coypu dataset. Note that a site random effect was not included here, as the spatial scale of the coypu dataset was limited. However, such effects may be important to consider in broader-scale programs where unobserved spatial heterogeneity is likely to be more pronounced, as in the muskrat case study.

\subsection{Muskrats in the Netherlands}\label{muskrats-in-the-netherlands}

\begin{figure}[H]

{\centering \includegraphics[width=0.7\linewidth]{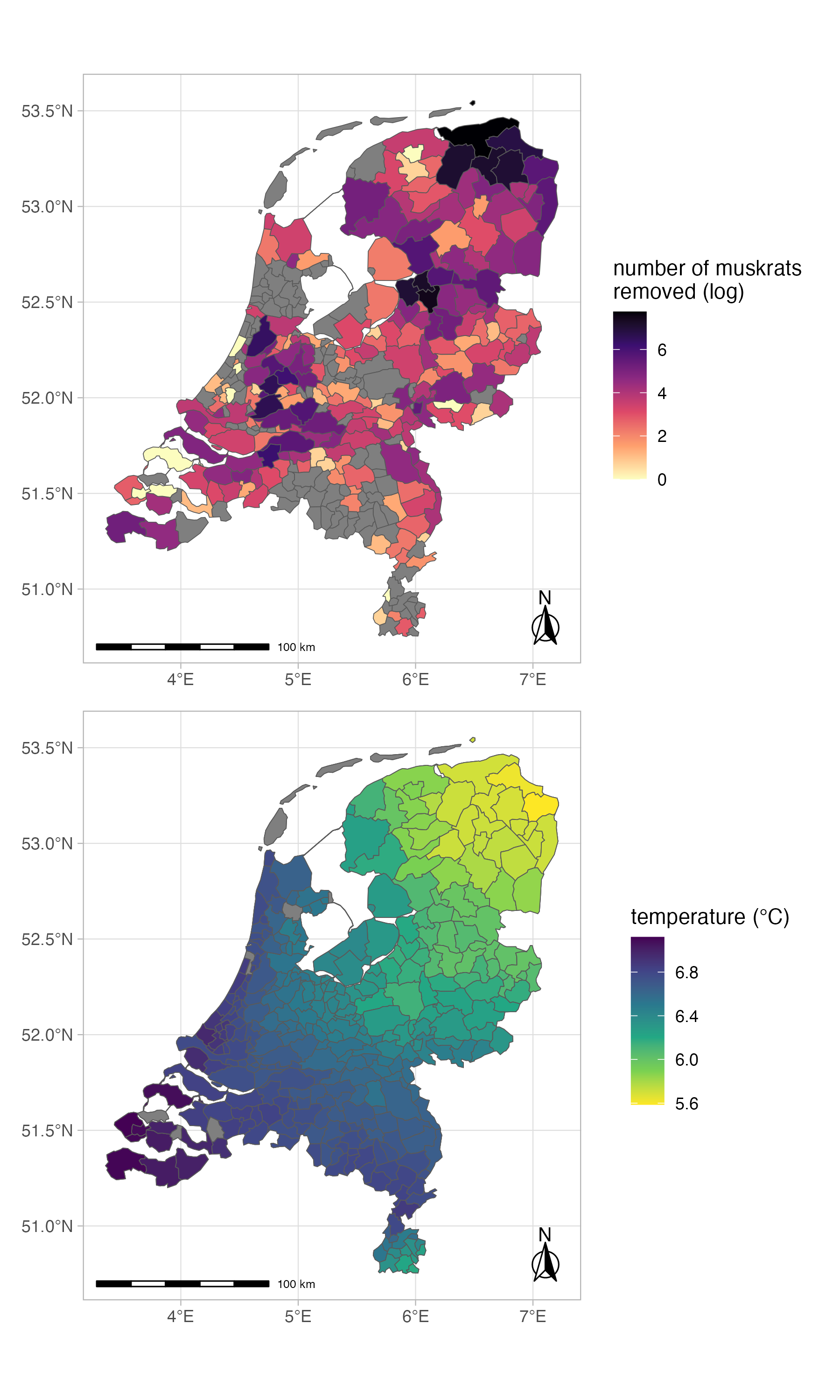} 

}

\caption{Total number of invasive muskrats removed over the period January-February-March (top panel), and the average 3-month temperature (bottom panel) across the Netherlands in 2014.}\label{fig:datsmuskrats}
\end{figure}

Removal data on muskrats in the Netherlands were collected by professional trappers. The data were registered in atlas blocks (5 x 5 km) per periods of four weeks. For this study, I focus on data from 2014, specifically from sampling occasions in January, February, and March. The data were made available through the LIFE MICA project (\citeproc{ref-Cartuyvels2024}{Cartuyvels et al. 2024}) and can be freely downloaded from \url{https://www.gbif.org/dataset/7d75109d-a6cb-4e90-89d0-79d08577c580} (\citeproc{ref-moerkens2021muskrat}{Moerkens et al. 2021}). The data, covering \(S = 215\) cities (out of the 342 cities in the Netherlands), are presented in Figure \ref{fig:datsmuskrats}. I fitted the same model as for the coypus data, except that I added a site random effect on abundance to accommodate the spatial variation that was not explained by temperature.

A key assumption underlying the proper use of multinomial N-mixture models is that of population closure, which assumes no births, deaths, immigration, or emigration occur during the trapping period. A straightforward approach to relax this assumption is to fit multiple years of data (a.k.a. stacking the data) into a standard multinomial N-mixture model. In this approach, year-site combinations are treated as separate sites, and year is included as a site covariate or random effect in the model. I used this method to evaluate a temporal effect on the relationship between temperature and abundance. Assuming an increase in temperature over time, one might predict a decoupling or weakening of the relationship between abundance and temperature. To test this, I conducted an additional analysis spanning the 1987--2014 period, modeling the slope of the temperature-abundance relationship as a linear function of time.

\subsection{Implementation}\label{implementation}

For all analyses, I used the statistical language \texttt{R} (\citeproc{ref-R_Core_Team_2024}{R Core Team 2024}). I used the \texttt{tidyverse} (\citeproc{ref-Wickham_2019}{Wickham et al. 2019}) suite of packages for data manipulation and visualization, \texttt{sf} (\citeproc{ref-Pebesma_2023}{Pebesma and Bivand 2023}) for dealing with spatial data and \texttt{krigR} (\citeproc{ref-Kusch_2022}{Kusch and Davy 2022}) to get temperature data. For the simulations, I used the \texttt{R} package \texttt{unmarked} (\citeproc{ref-Kellner_2023}{Kellner et al. 2023}), see the Simulation study section. For the two case studies, I fitted models within a Bayesian framework using Markov chain Monte Carlo (MCMC) algorithms. I used both the \texttt{NIMBLE} (\citeproc{ref-de_Valpine_2017}{de Valpine et al. 2017}) and the \texttt{ubms} (\citeproc{ref-Kellner_2021}{Kellner et al. 2021}) packages. The former offers high flexibility, enabling users to define custom likelihoods, though it requires manual coding, while the latter features simpler syntax with pre-built multinomial N-mixture models, albeit limited to a Poisson distribution for abundance. I specified weakly informative priors for all parameters, specifically normal distributions with mean 0 and standard deviation 1.5 for regression parameters, and a uniform distribution for the standard deviation of the random effects. I ran two chains for a total of 200,000 iterations with a burn-in of 20,000 iterations. I summarized posterior distributions with posterior mean and 95\% credible intervals. I assessed convergence using standard Bayesian diagnostics: the R-hat statistic (values close to 1 indicate convergence), effective sample size (which reflects the amount of independent information in the posterior sample, should be \textgreater{} 100), and visual inspection of trace plots (which should show good mixing and stationarity of the chains).

\section{Results and discussion}\label{results-and-discussion}

\begin{figure}[H]

{\centering \includegraphics[width=0.9\linewidth]{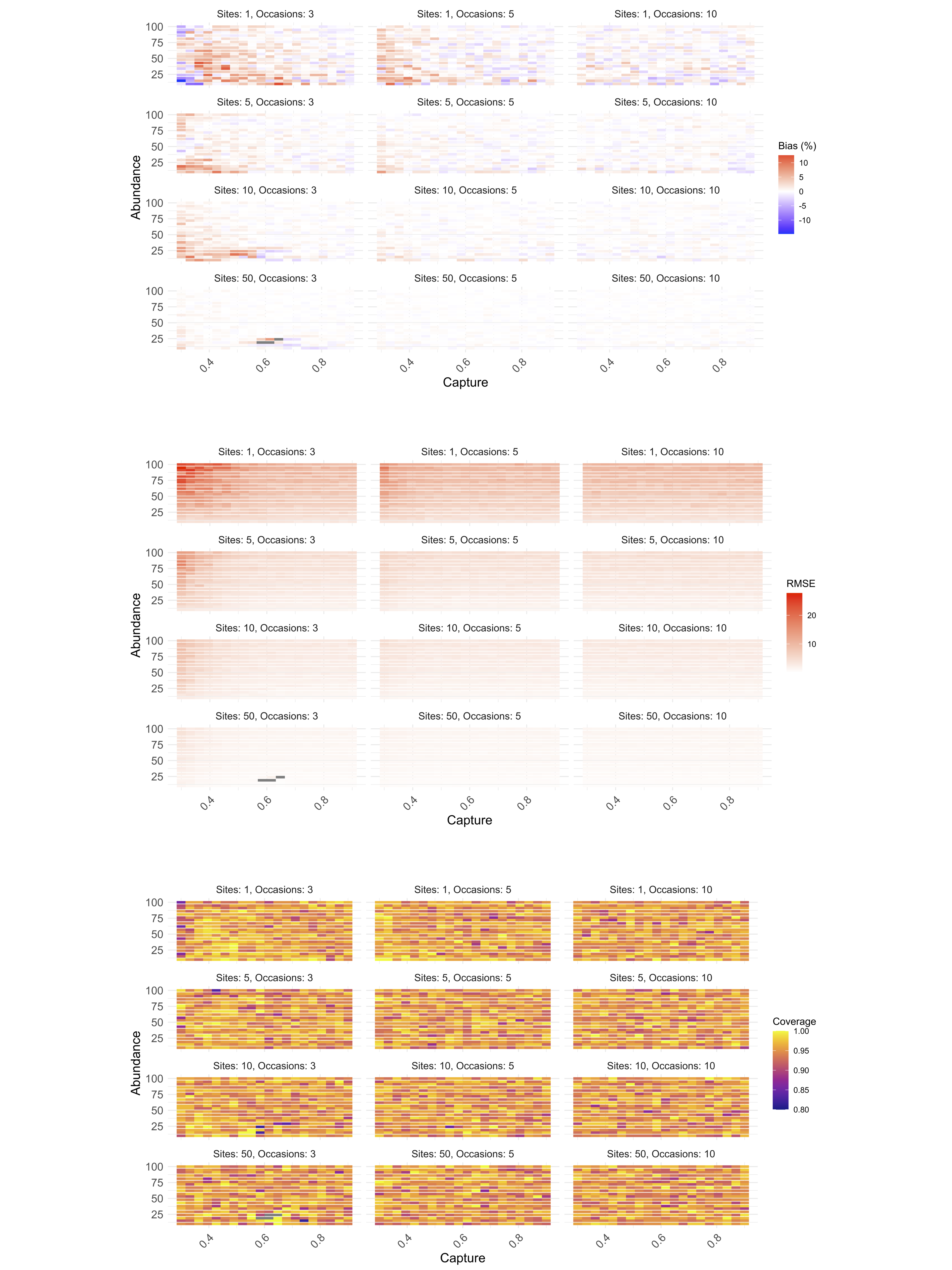} 

}

\caption{Relative bias (top panel), root mean square error (RMSE; middle panel) and coverage of the 95\% confidence interval (bottom panel) for abundance estimates from a multinomial N-mixture model with constant parameters. Capture probabilities (X-axis) range from 0.3 to 0.9, while abundance (Y-axis) varies between 10 and 100 individuals. Scenarios consider 3, 5, and 10 capture occasions (columns) and 1, 5, 10, and 50 sites (rows). Results are based on 100 simulations.}\label{fig:bias}
\end{figure}

\begin{figure}[H]

{\centering \includegraphics[width=0.75\linewidth]{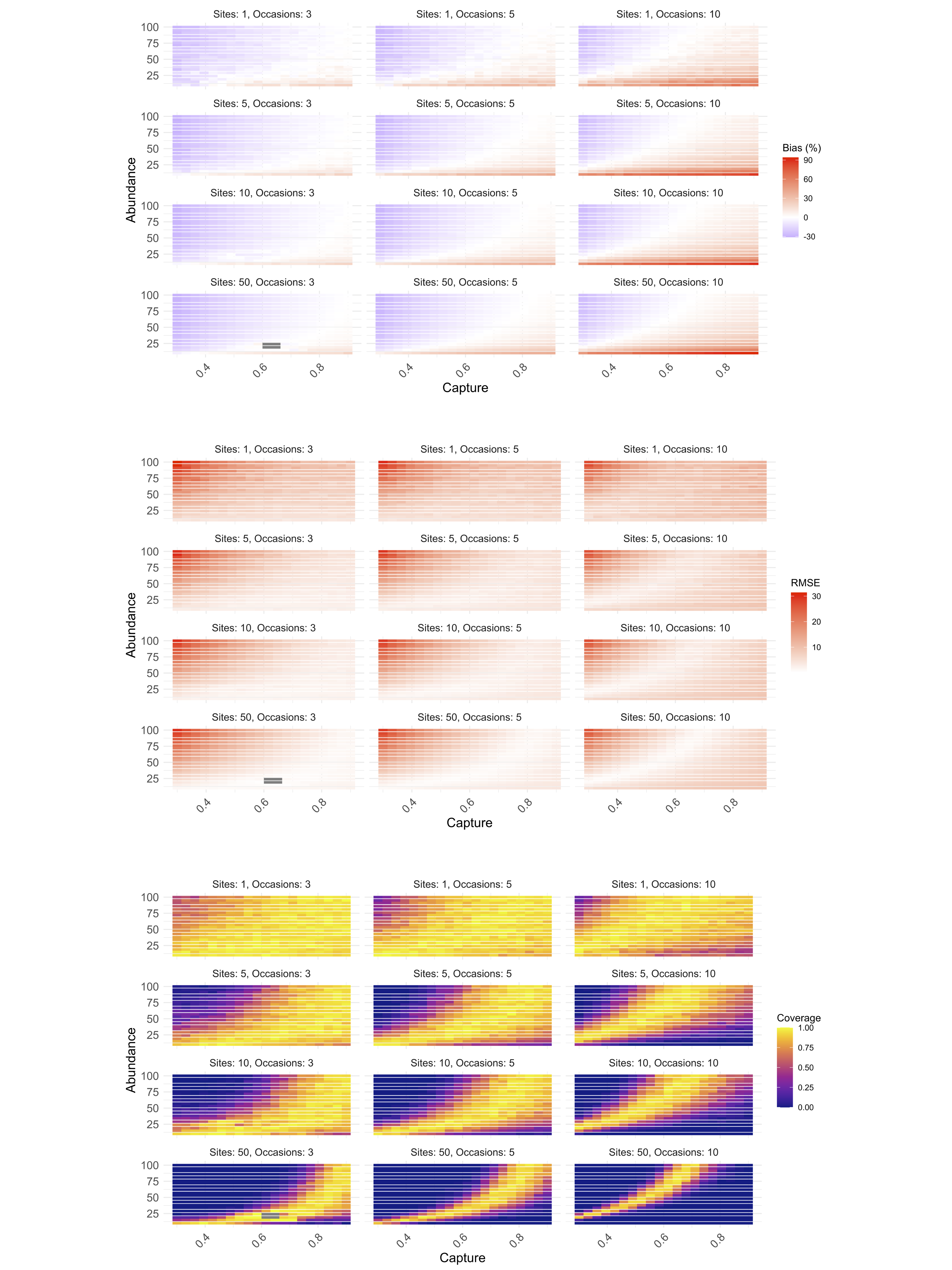} 

}

\caption{Relative bias (top panel), root mean square error (RMSE; middle panel) and coverage of the 95\% confidence interval (bottom panel) for abundance estimates from a multinomial N-mixture model with constant parameters, fitted to data where the closure assumption was deliberately violated. Between capture occasions, individuals remained in the population with probability 0.8, and new individuals arrived according to a Poisson process with mean 1, introducing both emigration and immigration between sampling events. Capture probabilities (X-axis) range from 0.3 to 0.9, while abundance (Y-axis) varies between 10 and 100 individuals. Scenarios consider 3, 5, and 10 capture occasions (columns) and 1, 5, 10, and 50 sites (rows). Results are based on 100 simulations.}\label{fig:biasclosure}
\end{figure}

The results of the simulation study are presented in Figures \ref{fig:bias} and \ref{fig:biasclosure}. Overall, the analysis revealed minimal bias, with the exception of one site (first row) that showed a notable deviation (Figure \ref{fig:bias}).

Increasing the number of sites to 10 significantly reduced this bias, and no bias was observed with 50 sites, supporting the recommendation by (\citeproc{ref-Dorazio2005}{Dorazio et al. 2005}) to analyze data jointly rather than separately.

When the closure assumption was not met, the analysis revealed that both bias and precision metrics were highly sensitive to the introduction of between-occasion population dynamics (Figure \ref{fig:biasclosure}). Specifically, relative bias increased and coverage dropped in many scenarios, particularly when detection probability was low or the number of sites and occasions was limited. These results highlight how violations of closure can substantially compromise the reliability of abundance estimates derived from closed-population models.

Overall, these findings align with previous simulation studies on binomial N-mixture models (e.g., \citeproc{ref-Womack2019}{Womack-Bulliner et al. 2019}, \citeproc{ref-Fogarty2021}{Fogarty and Fleishman 2021}).

To enhance reproducibility, I provide the code for the simulation study in the Supplementary Material. This resource can be adapted for various purposes, such as conducting custom simulation studies or designing removal protocols and conducting power analyses.

\begin{figure}[H]

{\centering \includegraphics[width=0.98\linewidth]{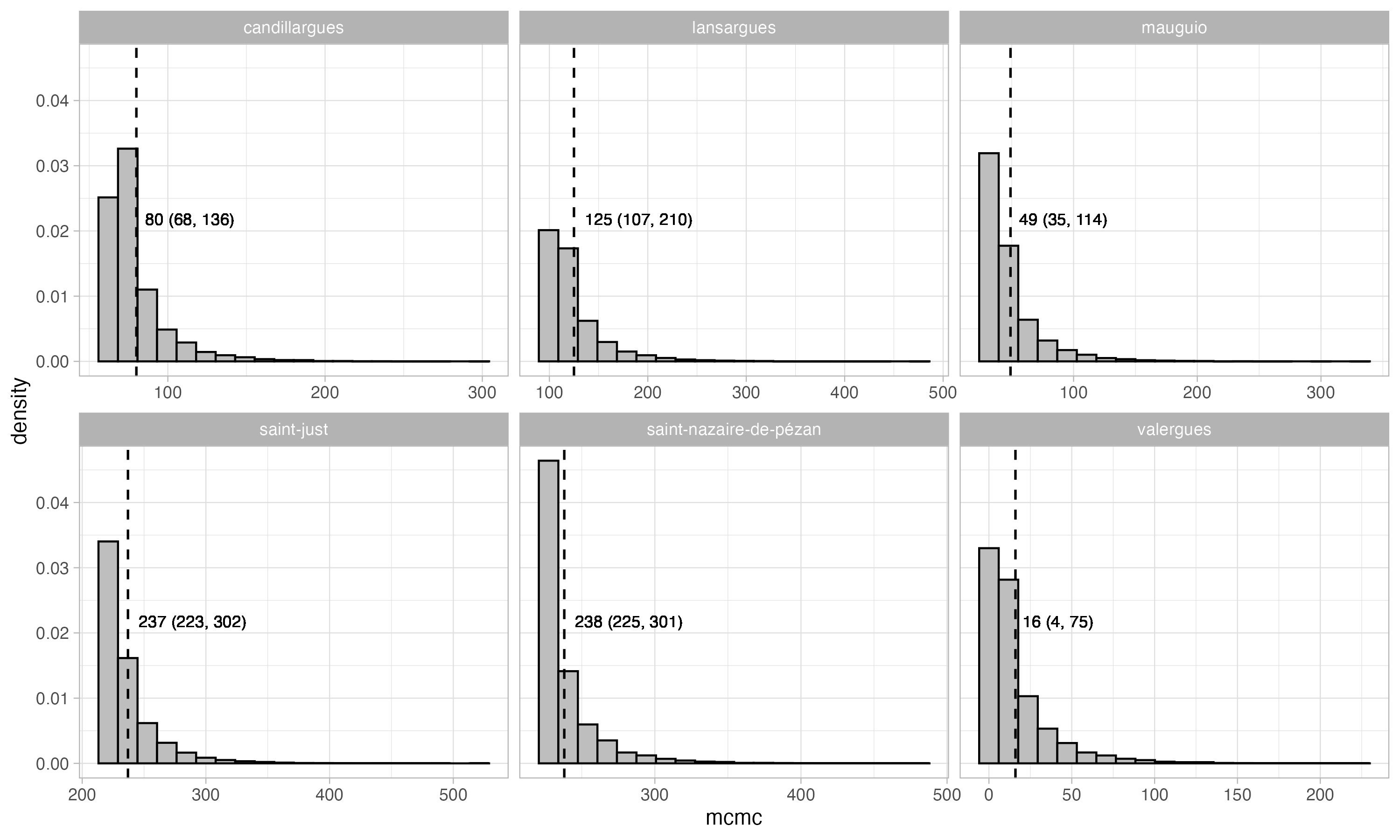} 

}

\caption{Posterior density plots for coypu abundance across several cities in the Hérault department, France, in 2022. The vertical shaded line indicates the posterior mean abundance, accompanied by its credible interval. See Table 1 for the raw data.}\label{fig:pdcoypus}
\end{figure}

In the coypus case study, temperature was found to have a negative effect on abundance, with a slope estimate of -0.14 (-0.22, -0.07). Capture probabilities were estimated at 0.43 (0.29, 0.49) in February, 0.35 (0.19, 0.44) in March and 0.84 (0.35, 1.00) in April. The posterior distributions of abundance across the different sites are presented in Figure \ref{fig:pdcoypus}.

The Poisson assumption does not appear valid based on the estimated abundance in Figure \ref{fig:pdcoypus}, which shows a skewed posterior distribution of site-level abundances and wide credible intervals, suggesting greater variability than expected under a Poisson distribution. This is confirmed by a goodness-of-fit test, which yielded a p-value of 0.04, indicating a lack of fit. Fortunately, this limitation can be addressed by relaxing the Poisson assumption and use a negative binomial distribution instead. This adjustment can be implemented in both \texttt{NIMBLE} and \texttt{unmarked} but is not currently supported by \texttt{ubms}. Interestingly, under the negative binomial model, the effect of temperature on abundance was no longer significant, with a slope estimate of -0.27 (-1.29, 0.60). Moreover, the goodness-of-fit test for the abundance component of the model no longer indicated a lack of fit, with a p-value of 0.39.

\begin{figure}[H]

{\centering \includegraphics[width=0.85\linewidth]{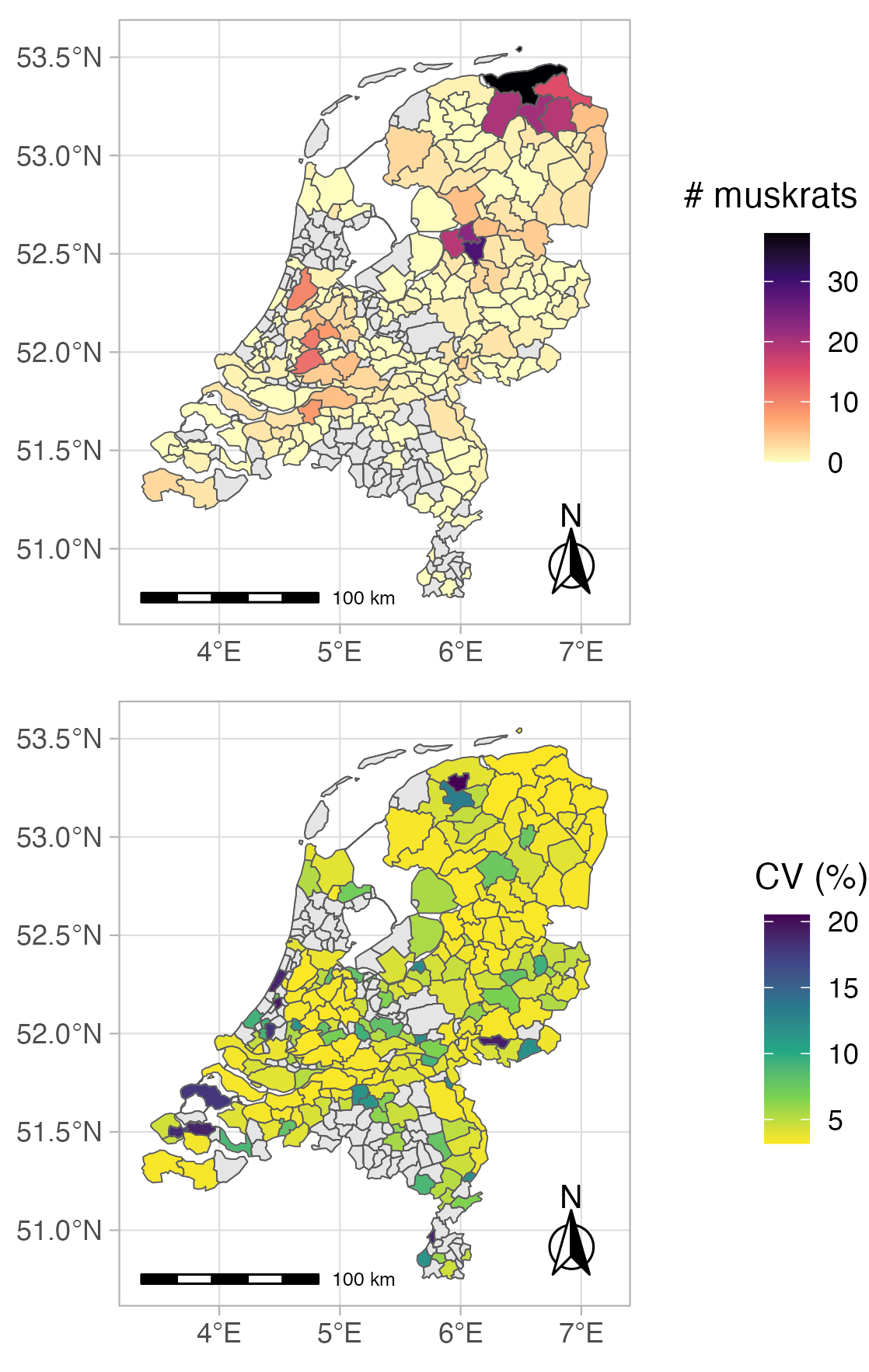} 

}

\caption{Posterior mean estimates of the number of muskrats remaining after removal in the Netherlands in 2022 (top panel) and the corresponding coefficient of variation (bottom panel). See Figure 1 for the raw data.}\label{fig:muskrats}
\end{figure}

In the muskrats case study, temperature was found to have a negative effect on abundance, with a slope estimate of -0.48 (-0.70, -0.26). The standard deviation of the site random effect was estimated at 1.62 (1.46, 1.79). Capture probabilities were estimated at 0.12 (0.05, 0.25) in January, 0.25 (0.15, 0.35) in February and 0.60 (0.53, 0.75) in March. Estimated abundance across sites after removal is presented in Figure \ref{fig:muskrats}. A goodness-of-fit test for the abundance component of the model yielded a p-value of 0.44, indicating no evidence of lack of fit. However, the test for the observation component revealed a lack of fit (p-value = 0), for reasons that remain to be investigated.

An important feature of multinomial N-mixture models is their ability to quantify uncertainty, which is often overlooked in spatial analyses. Here, I provide the coefficient of variation to represent the uncertainty surrounding abundance estimates. This metric can help identify specific areas where increased sampling effort might be beneficial to improve estimate precision.

Turning to the open-population analysis, the results revealed a negative trend in the slope, estimated at -0.040 (-0.042, -0.039), providing evidence to support the hypothesis of a temporal weakening of this relationship.

Several perspectives arise from this work. From a methodological standpoint, this study highlights the suitability of hierarchical models for capturing dependencies in space and time, which are common in ecological systems and removal experiments in particular. Two areas stand out as particularly worth exploring. First, although much of the paper focuses on static models under the assumption of population closure, I have taken a first step toward relaxing this assumption by fitting an open-population model in the muskrats case study. This extension does not yet incorporate explicit demographic mechanisms, but it illustrates the potential of such models to better account for temporal population dynamics. When the mechanisms underlying population dynamics---such as survival, recruitment, or dispersal---are of interest, multinomial N-mixture models can be extended to open populations (\citeproc{ref-Matechou2016}{Matechou et al. 2016}, \citeproc{ref-Link2018}{Link et al. 2018}, \citeproc{ref-Zhou2019}{Zhou et al. 2019}, \citeproc{ref-Tiberti2021}{Tiberti et al. 2021}). These extended models can be implemented using \texttt{unmarked} or \texttt{NIMBLE}, though they are not yet available in \texttt{ubms}.

A second area of investigation concerns the spatial dimension of multinomial N-mixture models, particularly the assumption of independence among removal sites (i.e., that removals at one site do not influence those at another). One possible solution is to include site random effects, as demonstrated in the muskrat case study. To better address spatial autocorrelation, restricted spatial regression (RSR) can also be employed (\citeproc{ref-Johnson2013}{Johnson et al. 2013}, \citeproc{ref-Broms2014}{Broms et al. 2014}) to impose a structure where spatially adjacent sites are modeled to have correlated random effects, effectively accounting for spatial autocorrelation. RSR models are advantageous because their random effects are constructed to be uncorrelated with fixed covariates, avoiding potential confounding issues, and they are computationally efficient. These models are easy to fit using \texttt{ubms} and can also be implemented in \texttt{NIMBLE} (\citeproc{ref-Cook2022}{Cook et al. 2022}), although they are not currently supported by \texttt{unmarked}. A promising extension would involve adapting the covariance structure in these models to account for stream networks (\citeproc{ref-Gimenez2024}{Gimenez 2024}, \citeproc{ref-Lu2024}{Lu et al. 2024}), which is particularly relevant for semi-aquatic rodents.

Third, in many real-world removal programs, detection probability is strongly influenced by effort-related factors such as trap density, frequency of checks, or personnel availability (e.g., \citeproc{ref-Davis2016}{Davis et al. 2016}). Unfortunately, explicit effort data were not available for the case studies analyzed here. To account for temporal variation in detection probability, likely driven in part by effort and other factors such as weather, seasonal activity, or habitat changes, I modelled detection as a time-varying parameter. While this approach provides a practical solution when effort is missing or inconsistently reported, it does not allow for prospective analyses, such as estimating the level of effort required to achieve a desired level of precision on abundance estimates or statistical power to detect changes or trends in abundance over time. Addressing such questions requires explicitly incorporating effort into the model. I therefore emphasize the importance of properly recording removal effort and caution against interpreting its omission here as a reason to neglect it. On the contrary, I hope this work underscores the value of systematic effort reporting to improve both inference and future survey design.

From an ecological perspective, the main contribution of this paper is to demonstrate the estimation of abundance for rodent populations in Europe. The European Union Regulation No.~1143/2014 was established to mitigate the negative impacts of invasive alien species on biodiversity. This regulation mandates measures to prevent the introduction of invasive alien species and manage their established populations. In this study, I focused on two species listed as species of Union Concern under the regulation, which requires member states to implement appropriate management actions.

In this context, although it is straightforward to calculate the number of coypus or muskrats remaining after removal campaigns (including associated uncertainty; see Figure \ref{fig:muskrats}), abundance estimates realize their full potential only when integrated into a decision-making framework. Such a framework allows for evaluating the efficiency of eradication or control efforts, optimizing the allocation of limited resources, and accounting for uncertainty in the management of invasive species (\citeproc{ref-Shea2002}{Shea et al. 2002}, \citeproc{ref-Adams2024}{Yackel Adams et al. 2024}).

\section{Acknowledgments}\label{acknowledgments}

I would like to warmly thank Nathalie Vazzoler-Antoine and Tanguy Lebrun for sharing the coypus data, and the team LIFE MICA for sharing the muskrats data. This work was funded by a grant from the University of Montpellier through its ExposUM institute.

\subsection{Data availability statement}\label{data-availability-statement}

Data and code are available at \href{https://github.com/oliviergimenez/counting-rodents}{https://github.com/oliviergimenez/counting-rodents}.

\section{References}\label{references}

\phantomsection\label{refs}

\end{document}